\documentclass[runningheads,a4paper]{cmmr2023}
\usepackage[style=lncs]{biblatex}
\usepackage{times}
\usepackage{amssymb}
\usepackage{graphicx}
\usepackage{subfig}
\usepackage{float}
\usepackage{booktabs}
\usepackage{url}

\newcommand{\keywords}[1]{\par\addvspace\baselineskip
\noindent\keywordname\enspace\ignorespaces#1}
\begin{document}

\mainmatter  

\title{JAZZVAR: A Dataset of Variations found within Solo Piano Performances of Jazz Standards for Music Overpainting}

\titlerunning{JAZZVAR}


\author{Eleanor Row\inst{1} \and Jingjing Tang\inst{1} \and Gy\"orgy Fazekas\inst{1} \thanks{This work is supported by the UKRI Centre for Doctoral Training in Artificial Intelligence and Music, funded by UK Research and Innovation [grant number EP/S022694/1]. J.Tang is a research student also supported jointly by the China Scholarship Council and Queen Mary University of London. Special thanks to Max Graf, for his help in creating the GUI, and to both him, Huan Zhang, and Corey Ford for reviewing our paper.}}
%

\institute{Centre for Digital Music, Queen Mary University of London\\\email{e.r.v.row@qmul.ac.uk}}


\maketitle


\begin{abstract}
Jazz pianists often uniquely interpret jazz standards. Passages from these interpretations can be viewed as sections of variation. We manually extracted such variations from solo jazz piano performances. The JAZZVAR dataset is a collection of 502 pairs of `\textit{Original}' and `\textit{Variation}' MIDI segments. Each \textit{Variation} in the dataset is accompanied by a corresponding \textit{Original} segment containing the melody and chords from the original jazz standard. Our approach differs from many existing jazz datasets in the music information retrieval (MIR) community, which often focus on improvisation sections within jazz performances. In this paper, we outline the curation process for obtaining and sorting the repertoire, the pipeline for creating the \textit{Original} and \textit{Variation} pairs, and our analysis of the dataset. We also introduce a new generative music task, Music Overpainting, and present a baseline Transformer model trained on the JAZZVAR dataset for this task. Other potential applications of our dataset include expressive performance analysis and performer identification.

\end{abstract}

\section{Introduction}\label{sec:intro}

The growing interest in generative music models has led to the exploration of their potential in specialised music composition tasks. As current trends often focus on generating complete songs or music continuation tasks \cite{briotArtificialNeuralNetworks2021, briotDeepLearningMusic2020}, there is a lack of datasets designed for specialised music tasks. However, these specialised music tasks, such as music infilling \cite{patiLearningTraverseLatent2019, 9747817} and composition style transfer \cite{Mukherjee2021ComposeInStyleMC, Wu2021StyleFormerRA}, could contribute to the development of artificial intelligence (AI) tools in music composition.

We introduce Music Overpainting as a novel specialised generative music task, inspired by the concept of overpainting in fine art and Liszt's compositional approaches to rearrangement in his piano transcriptions from classical music. Music Overpainting generates variations by providing a rearrangement of a music segment. While the task aims to reframe the musical context by changing elements such as rhythmic, harmonic, and melodic complexity and ornamentation, the core melodic and harmonic structure of the music segment is preserved. Compared to related music generation tasks such as compositional style transfer \cite{daiMusicStyleTransfer2018} and music infilling \cite{patiLearningTraverseLatent2019, 9747817}, Music Overpainting creates small variations within the same style and retains perceptible similarities in the underlying melodic contour and harmonic structure of the music segment. Outputs from Music Overpainting could be used in AI tools for music composition, to add variation and novelty to desired sections of music. 

Our motivation for creating this dataset stems from the lack of available datasets for novel and specialised generative music tasks. Not only did we find that there was a lack of clean and high-quality MIDI data for investigating tasks such as Music Overpainting, but also in the context of solo jazz piano music in general. Most existing jazz datasets consist of transcriptions of improvised ``solo'' sections within a jazz performance or feature multiple instruments. Few datasets feature interpretations of the ``head'' section, containing the main musical theme, for solo piano only. Additionally, we found that many jazz datasets do not include performances from female musicians, so we are proud to include several extracts of performances from female jazz pianists within our dataset. Our dataset helps to fill this gap, while also providing insights into how jazz pianists rearrange standards for solo piano from a music information retrieval (MIR) perspective. 
\vspace*{-8mm}
\begin{table}[!hbt]
\caption{Overview of \textit{Original} and \textit{Variation} Segments.}
\label{tab:definitions}
\centering
\begin{tabular}{p{0.3\linewidth}p{0.3\linewidth}p{0.3\linewidth}}
\toprule
\textbf{Feature} & \textbf{Original} & \textbf{Variation} \\
\hline
Segment length & 4 bars& misc. \\
Location & ``head'' section & ``head'' section \\ 
File format & Manually-transcribed MIDI & Automatically-transcribed MIDI \\
Musical format & Melody and chords & Two-handed solo piano \\
Type & Lead sheet of &  Piano performance of \\
 & jazz standard &  jazz standard \\
Source & MuseScore & Youtube \\
\bottomrule
\end{tabular}
\vspace*{-2mm}
\end{table}

\begin{figure}[!hbt]
\centering
\includegraphics[width=\textwidth]{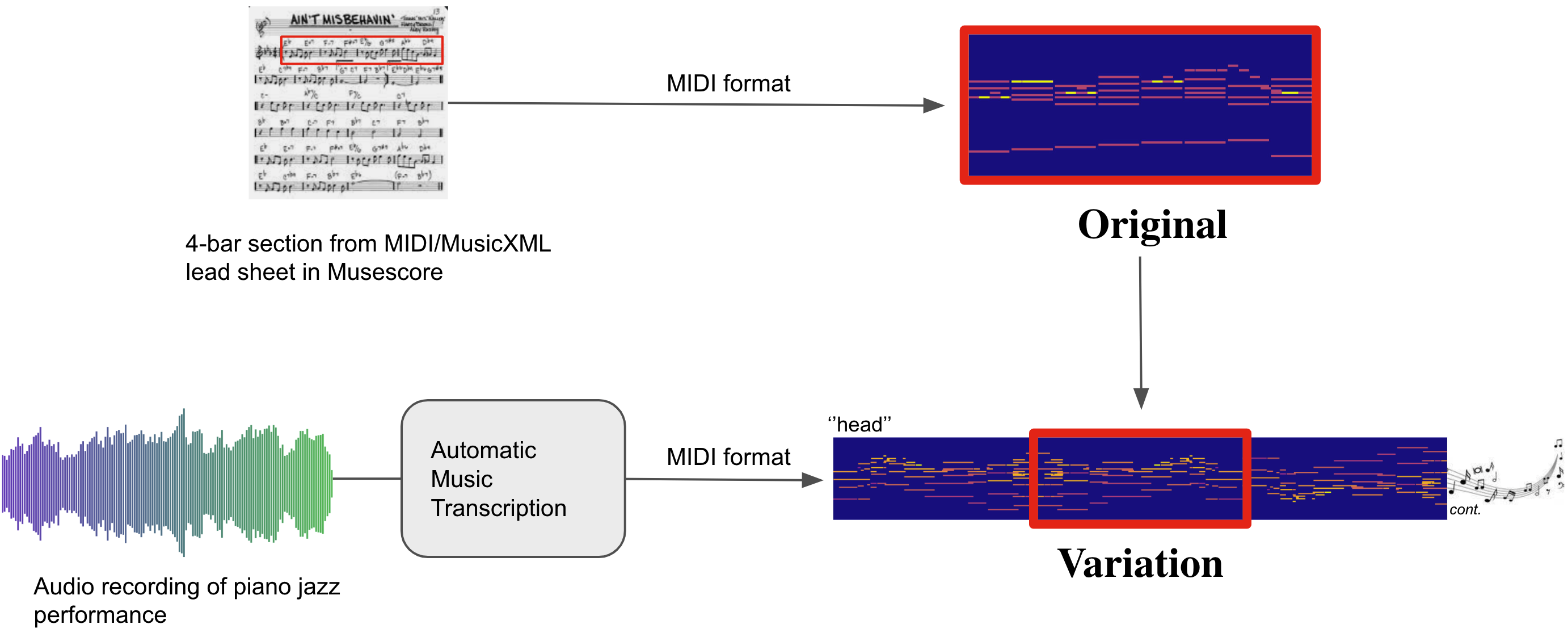}
\caption{The process of creating \textit{Original} and \textit{Variation} pairs. \textit{Original} sections are MIDI segments from a lead sheet transcription of a jazz standard. Audio of a piano performance playing the same jazz standard is transcribed automatically into MIDI. A \textit{Variation} is found by manually searching for passages that are melodically and harmonically similar to the \textit{Original} in the ``head'' section of the piano performance.}
\label{fig:pipeline}
\vspace*{-5mm}
\end{figure}

The JAZZVAR dataset comprises of 502 pairs of \textit{Original} and \textit{Variation} MIDI segments from 22 jazz standards, 47 performances, and 35 pianists. An \textit{Original} segment is 4-bars long and manually transcribed from a lead sheet of a jazz standard. A \textit{Variation} segment is manually found from an automatically transcribed piano performance of the same jazz standard. We find \textit{Variation} segments by searching for passages that are melodically and harmonically similar to \textit{Original} segments. Figure \ref{fig:pipeline} shows more details of the data curation pipeline. Table \ref{tab:definitions} provides more information about the \textit{Original} and \textit{Variation} segments. The jazz standards and the piano performances in our dataset are under copyright, therefore the JAZZVAR dataset cannot currently be made available for direct download. However, researchers will be allowed to access the dataset on request.

The JAZZVAR dataset serves as a foundation for exploring the Music Overpainting task across genres. What we refer to as \textit{Variations} are passages of music from a jazz standard that have been reinterpreted or rearranged by jazz pianists'. However, we can view these reinterpretations as variations on the melody and chords of the jazz standards. We use the \textit{Original} and \textit{Variation} pairs in the dataset to train a Music Transformer model to generate novel passages of variation from a simple MIDI primer. By presenting this novel dataset and introducing the Music Overpainting task, we aim to contribute to the field of generative music research and encourage further exploration of the relationship between composers and AI tools in various music genres.

The remainder of this paper is organised as follows: Section \ref{related} provides an overview of related datasets in the field of generative music and MIR, Sections \ref{dataset} and \ref{analysis} present an in-depth description and analysis of the JAZZVAR dataset, Section \ref{Task} introduces Music Overpainting as a generative music task and uses the JAZZVAR dataset to train the Music Transformer model for generation.

\section{Related Works} \label{related}
Existing jazz datasets that can be used for MIR and Generative Music tasks often feature the improvisation or solo section only of the jazz performance. The Weimar Jazz Database (WDB) \cite{Pfleiderer:2017:BOOK}, consists of 456 manually transcribed solos by 78 performers and contains no solo piano performances. The DTL1000 dataset \cite{dixon2022history} from the “Dig That Lick” project is a set of 1750 automatically transcribed solos from 1060 tracks. However, it is not clear how many of these tracks are piano solo tracks. 

The Million Song Dataset (MSD) \cite{Bertin-Mahieux2011} is a collection of audio features and metadata for one million contemporary popular music tracks. While the MSD does not specifically focus on jazz, it does include a substantial number of jazz recordings that could be used for comparative analysis. The Lakh MIDI Dataset (LMD) \cite{Manilow2019CuttingMS} is a collection of 176,581 unique MIDI files that are matched to songs within the Million Song Dataset using Dynamic Time Warping-based alignment methods \cite{7471641}. Similarly, to the DTL1000 dataset, the MSD and the LMD have no specific focus on solo jazz piano performances.  

\section{JAZZVAR Dataset} \label{dataset}
\subsection{Data Collection}
\subsubsection{Repertoire}

A jazz standard is a well-known, and commonly played song in the jazz repertoire. Many popular songs composed in the early to mid-twentieth century for film, television, and musical theatre are now prominent jazz standards. Some of the more famous jazz standards include Gershwin's ``Summertime'' for the opera \textit{Porgy and Bess} (1935) and ``All the Things You Are'' by Jerome Kern and Oscar Hammerstein II for the musical \textit{Very Warm for May} (1939). These popular songs have been continually played and rearranged by jazz musicians for decades. Popular songs originating from these times contain a ``refrain'' section, which was the main theme of the song. In jazz music, the ``head'' section is often synonymous with these ``refrain'' sections. Many jazz musicians would learn the songs by ear, or through unofficial lead sheets, such as the ones circulated within the \textit{Fake Real Book}. Some jazz musicians, such as the trumpeter Miles Davis (1926-1991) and Thelonious Monk (1917-1982), composed music themselves and these pieces have also become famous jazz standards. 

Within this context, our goal was to find lead sheets of jazz standards and audio recordings of solo piano performances of jazz standards. The first publication dates of the jazz standards in our dataset range between 1918 and 1966, while the performances span from the mid-twentieth to the beginning of the twenty-first century. 

\subsubsection{Jazz Standard Lead Sheets}

Lead sheets are condensed versions of song compositions that musicians have transcribed and passed through the community. They are presented as a single melodic line with accompanying chords.

We sourced MIDI and MusicXML lead sheets from MuseScore, created by users who often referenced the \textit{Fake Real Book}. Candidate pieces were found using the following criteria:
\begin{enumerate}
    \item entirely in 4/4 timing,
    \item jazz standards mostly consisting of popular songs from the early to mid-twentieth century.
\end{enumerate}

The lead sheets were cleaned and corrected by removing introductions and verses, to retain only the refrain section. Songs with repeated refrains were further edited to include only the final repeat. We converted any MusicXML files to MIDI and made corrections by referencing the chords in lead sheets. In some cases, we transcribe the chords and melody by ear from early recordings of popular songs or completely rewrite the MIDI, as many of the source files were corrupt. In total, we collected and cleaned 234 jazz standards, of which a subset of 22 appear within the JAZZVAR dataset. 

\subsubsection{Audio of Jazz Solo Piano Performances}

To compile a list of solo piano performances of jazz standards, we manually searched for well-known jazz pianists' solo performances on Spotify and Youtube that matched the list of 234 MIDI lead sheets we had collected. We also used the \textit{Solo piano jazz albums}\footnote{See Wikipedia: \url{https://en.wikipedia.org/wiki/Category:Solo_piano_jazz_albums}} category on Wikipedia to help find performances. We gathered Spotify Metadata for these performances, which we used to collect the respective audio data. This approach allowed us to compile a diverse set of performances, including some by female pianists, and to capture the rich history of jazz piano performance.
\subsection{Automatic Music Transcription of Jazz Audio} \label{AMT}

Automatic Music Transcription (AMT) algorithms such as  \cite{kongGiantMIDIPianoLargeScaleMIDI2022, Hawthorne2021SequencetoSequencePT} enable us to transcribe audio recordings into MIDI representations. According to results from a listening test conducted by Zhang et al.\ \cite{zhang2022atepp}, the High-Resolution transcription system proposed by Kong et al.\ \cite{kongGiantMIDIPianoLargeScaleMIDI2022} is preferred over the other two systems by participants in terms of conserving the expressiveness of the performances. We used the Spotify metadata to download the jazz audio from Youtube and applied the High-Resolution model \cite{kongGiantMIDIPianoLargeScaleMIDI2022} to transcribe the downloaded jazz audios into MIDIs. In total, we collected and transcribed 760 audio recordings covering a wide range of performances from 148 albums by 101 jazz pianists, of which a subset of 47 performances appear within the JAZZVAR dataset. 

\subsection{Pair Matching Process} \label{PMP}

We segmented 4 bar sections from the MIDI lead sheets by taking into consideration the phrases in the main melody. As the jazz standards that we chose were all in 4/4 time, most of the phrases were contained within a 4-bar structure. We labeled these four bar sections as \textit{Original} segments. We segmented 22 jazz standards and collected an average of 6 segments per standard. In order to create our \textit{Variation} segments to form a data pair, we manually searched through the AMT solo jazz piano performances of the jazz standards and found segments that were melodically and harmonically similar to the \textit{Original} segment for each jazz standard. To facilitate the matching process for finding \textit{Original} and \textit{Variation} pairs, we created a Python application with a graphical user interface (GUI), which allowed us to view and listen to individual \textit{Original} segments. \footnote{We plan to release the GUI for reproducing our dataset. A GitHub page will be released by the publication of the paper.} We then searched through the AMT jazz performances and saved passages that closely corresponded to the \textit{Original} segments melodically and harmonically.

\section{Analysis} \label{analysis}

\subsection{Experimental Dataset Analysis}

We calculated several musical statistics across the dataset to provide insights into the dataset's musical content and structure according to \cite{Dong2018MuseGANMS}. We compared the differences between the \textit{Original} and the \textit{Variation} sections and summarise several characteristic features in Table \ref{tab:results}.
\begin{table}[!hbt]
\vspace*{-6mm}
\caption{Means and standard deviations for various statistics for combined segments in \textit{Original} and \textit{Variation} sections.}
\label{tab:results}
\centering
\begin{tabular}{l@{\hskip 0.1in}cc@{\hskip 0.1in}cc}
\toprule
\textbf{Feature} & \multicolumn{2}{c}{\textbf{Originals}} & \multicolumn{2}{c}{\textbf{Variations}} \\
 & Mean & SD & Mean & SD \\
\hline
Pitch Class Entropy & 2.94 & 0.24 & 3.13 & 0.24 \\
Pitch Range & 36.44 & 3.60 & 47.20 & 10.91 \\
Polyphony & 5.30 & 0.28 & 5.01 & 2.08 \\
Number of Pitches & 16.08 & 0.28 & 29.42 & 8.05 \\
Pitch in Scale & 0.89 & 0.24 & 0.83 & 0.08 \\
\hline
\end{tabular}
\vspace*{-5mm}
\end{table}

\subsubsection{Pitch Class Entropy} The higher mean pitch class entropy in the \textit{Variation} segments (3.13) compared to the \textit{Original} segments (2.94) suggests that jazz pianists tend to introduce more diversity in pitch distribution when interpreting jazz standards. This increased complexity and unpredictability in the variations reflect the improvisational and creative nature of jazz music.
\vspace*{-5.7mm}
\subsubsection{Pitch Range} The mean pitch range in the \textit{Variation} segments (47.20) is considerably larger than in the \textit{Original} segments (36.44), indicating that jazz pianists often expand beyond the range of pitches used within a jazz standard. This expanded pitch range could contribute to a richer and more expressive musical experience in the variations.
\vspace*{-5.7mm}
\subsubsection{Polyphony} Polyphony is defined as the mean number of pitches played simultaneously, evaluated only at time steps where at least one pitch is played. The mean polyphony is slightly lower in the \textit{Variation} segments (5.01) compared to the \textit{Original} segments (5.30). This suggests that jazz pianists may use fewer simultaneous pitches on average in their reinterpretations. However, the higher standard deviation in the \textit{Variation} segments (2.08) indicates that the polyphonic structures in these reinterpretations can be quite diverse.
\vspace*{-5.7mm}
\subsubsection{Number of Pitches} The higher mean number of pitches in the \textit{Variation} segments (29.42) compared to the \textit{Original} segments (16.08) implies that jazz pianists tend to incorporate more distinct pitches when rearranging jazz standards. This increase in the number of pitches adds to the complexity and expressiveness of the variations.
\vspace*{-5.7mm}
\subsubsection{Pitch in Scale} Pitch-in-scale rate is defined as the ratio of the number of notes in a certain scale to the total number of notes \cite{Dong2018MuseGANMS}. The slightly lower mean value of pitch in scale in the \textit{Variation} segments (0.83) compared to the \textit{Original} segments (0.89) indicates that jazz pianists may be more inclined to use pitches outside the underlying scale in their reinterpretations. This tendency could contribute to a more adventurous and explorative musical experience in the variations.

\vspace{3mm}
\noindent In summary, the analysis of the JAZZVAR dataset reveals that jazz pianists often introduce greater complexity, diversity, and expressiveness when rearranging jazz standards for solo piano.  Our findings highlight the dataset's potential for application in tasks such as Music Overpainting. Not only are these insights valuable for the development of specialised generative music models, but they also provide a better understanding of the creative process in jazz music.

\subsection{Comparison of Multiple Pianists}

Some of the jazz standards featured within the dataset are performed by multiple pianists. Therefore, there are some \textit{Original} segments that are matched to multiple \textit{Variation} segments from different pianists. To further highlight the diversity of variations within the dataset, we present a musical analysis of multiple pianists' interpretations of the same \textit{Original} segment, from the jazz standard ``All the Things You Are". 

\subsubsection{Melody}

The melody from the \textit{Original} segment was found and isolated within each \textit{Variation} segment. To obtain accurate representations of the melodies, we manually extracted the melody lines from the \textit{Variation} segments. This manual extraction process involved listening closely to the melody in the \textit{Original} in order to carefully isolate the melody line within the performances note by note, ensuring higher accuracy and fidelity of melodic extraction in comparison to an automatic approach. We then compared the isolated melodies to find their pitch and duration deviation from the ground truth, the melody from the \textit{Original} segment. We applied the Needleman-Wunsch \cite{GOTOH1982705, Kranenburg2009MusicalMF} alignment algorithm which aligns melodies by minimizing the differences in pitch class and duration between the corresponding notes. Based on the alignment results, we calculate the average deviation score using the following equation:
\begin{equation} \label{equ:average deviatoin}
    Average\ Deviation =  \frac{1}{n}\sum_{i=1}^{n} (PC_{i} + D_{i}),
\end{equation}
where $PC_{i}$ denotes the deviation of pitch class, $D_{i}$ denotes the deviation of note duration, and $i$ refers to the $i$-th note in the melody. We excluded the missing notes in the summation over the note sequences.

This average deviation score provides a measure of how similar the two melodies are, with lower scores indicating higher similarity. The deviation scores of the pianists' \textit{Variation} from the \textit{Original} melody can be found in Table \ref{tab:deviation}. Our results show that different pianists' have unique and individual approaches to interpreting the \textit{Original} melody. Some pianists, such as Leslie North, have a closer adherence to the \textit{Original} melody, while others, like Bill Evans, exhibit greater differences.

\begin{table}[H]
\vspace*{-5mm}
\caption{Average deviation from \textit{Original} melody for different pianists}
\label{tab:deviation}
\centering
\begin{tabular}{l@{\hskip 0.1in}c}
\toprule
\textbf{Pianist} & \textbf{Average Deviation} \\
\hline
Jim McNeely & 1.60 \\
McCoy Tyner & 1.04 \\
Roland Hanna & 1.50 \\
Lennie Tristiano & 0.92 \\
Elmo Hope & 1.11 \\
Leslie North & 0.65 \\
Bill Evans & 2.68\\
\bottomrule
\end{tabular}
\vspace*{-5mm}
\end{table}

\begin{figure}[H]
\vspace*{-5mm}
\centering
\includegraphics[width=\textwidth]{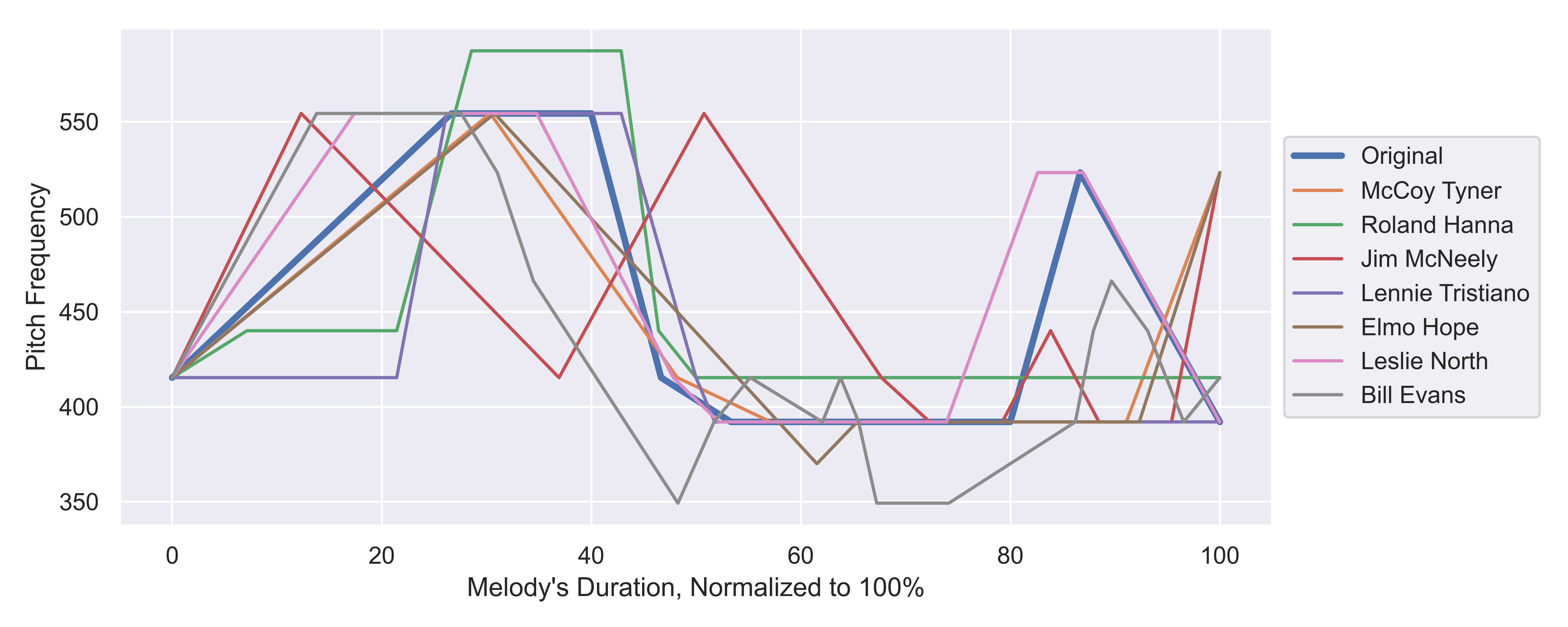}
\caption{The melodic contours of the melody taken from the jazz standard ``All The Things You Are'' (in Blue) and pianists' interpretations.}
\label{fig:contours}
\vspace*{-5mm}
\end{figure}

We also mapped the melodic contours of the performances to further explore the differences between the interpretations, using the Contourviz\footnote{Contourviz can be found in: \url{https://github.com/cjwit/contourviz}} package as shown in Figure \ref{fig:contours}. The visual representation of melodic contours allowed us to observe the overall structure and direction of the melody as it evolved throughout the performance. By comparing the melodic contours of different pianists, we found that some tended to be more experimental with their melodic choices, while others adhered more closely to the \textit{Original} melody. This variation in melodic contours provides additional evidence of the rich diversity present in our dataset. 

\subsubsection{Harmony}
\begin{figure}[!hbt]
\vspace*{-3mm}
\centering
\includegraphics[width=\textwidth]{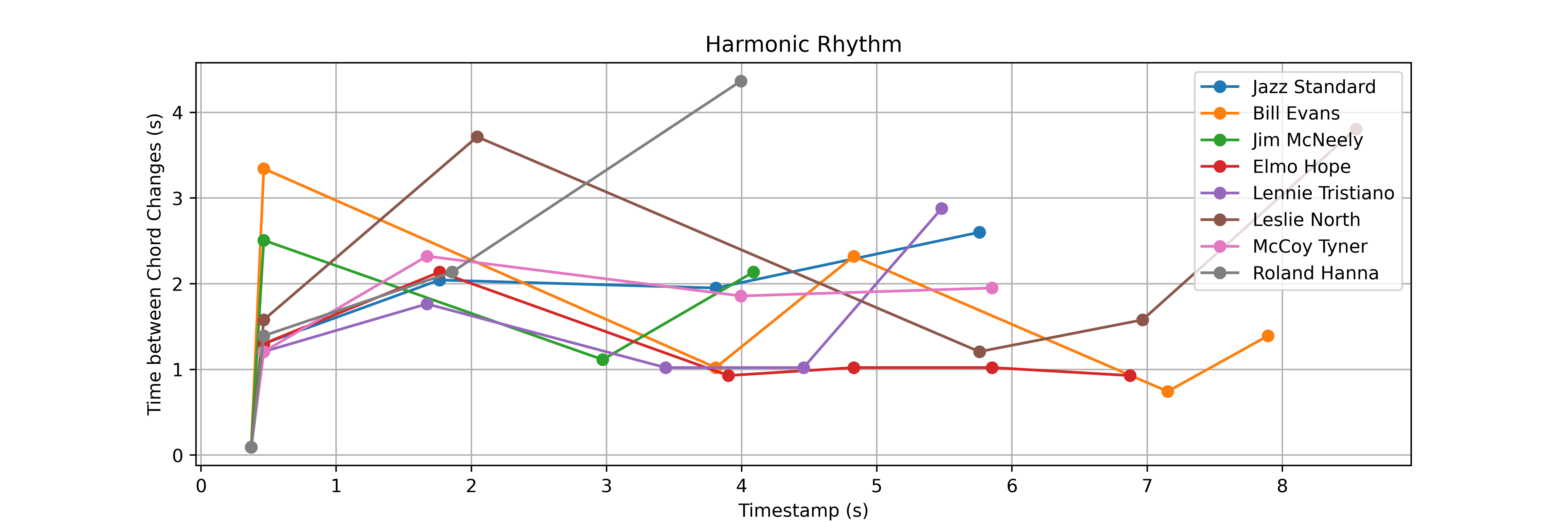}
\caption{A line graph comparison of the Harmonic Rhythm of the original melody (in Blue) and pianists' interpretations of the melody.}
\label{fig:harmonic_rhythm}
\vspace*{-3mm}
\end{figure}

The harmonies used within a performance can greatly impact the direction of the music and also the intention of the performer. To analyse some of the harmonic aspects of the dataset, we used Chordino and NNLS chroma \cite{Mauch2010ApproximateNT}. We set out to find the rate of harmonic change across each performance. As shown in Figure \ref{fig:harmonic_rhythm}, we found that some pianists had a higher harmonic rhythm (the rate of chord changes in a chord progression) than others. Other pianists added more chords to the chord progression, which sped up the harmonic rhythm. We observed that most pianists played in the key of the \textit{Original}, however, some transposed keys. Some pianists used the same chord progression as the \textit{Original} but altered specific chords. For example, Jim McNeely used a similar chord progression to the \textit{Original}, but modified a minor chord to major, resulting in a significant shift in the performance's intention and musical direction. We also observed that certain pianists used extended chords more extensively than others who played more closely to the \textit{Original}. Other pianists added more chords to the chord progression, which sped up the harmonic rhythm.

\vspace*{3mm}
\noindent Our analysis shows that the dataset contains a diverse range of interpretations, even when playing the same jazz standard. Within jazz, performers are individualistic and can be creative with their musical choices. The differences in melodic deviations, melodic contours, and harmonic rhythms between performances not only demonstrate the artistic freedom of each pianist but also indicates that the dataset could be a useful resource for those interested in expressive performance analysis or performer identification tasks. 

\section{Music Overpainting} \label{Task}

\subsection{Problem Definition}

As defined in Section \ref{sec:intro}, Music Overpainting is a generative music task that aims to create variations on pre-existing music sections. Within the context of the JAZZVAR dataset, we can specifically define the task as generating a \textit{Variation} segment from a given \textit{Original} segment. Given an \textit{Original} jazz standard segment $O$ from the JAZZVAR dataset, and a \textit{Variation} segment $V$, the goal of the Music Overpainting task is to find a reinterpretation $I(O)$ such that:
\begin{equation}
V = I(O)
\end{equation}

\subsection{Generation with Music Transformer} 
Transformers have been widely applied to generate music in genres such as Pop, Classical, as well as Jazz \cite{Huang2020POP, huang2018music, Wu2020TheJT}. Their convincing output demonstrate their capability of modeling musical structures and patterns. In this work, we adopted the design of Music Transformer \cite{huang2018music} which uses music motifs as primers for conditional generation. To train the transformer model, we concatenated the \textit{Variation} segments to the end of the \textit{Original} segments for each pair in the JAZZVAR dataset. In total, we obtained 502 concatenations and used 90\% for training and 10\% for validation. For the inference process, we treated the \textit{Original} segment as a primer and generated a \textit{Variation} segment following the probability distribution learned by the transformer model.

\begin{figure}[!hbt]
\centering
\includegraphics[width=\linewidth]{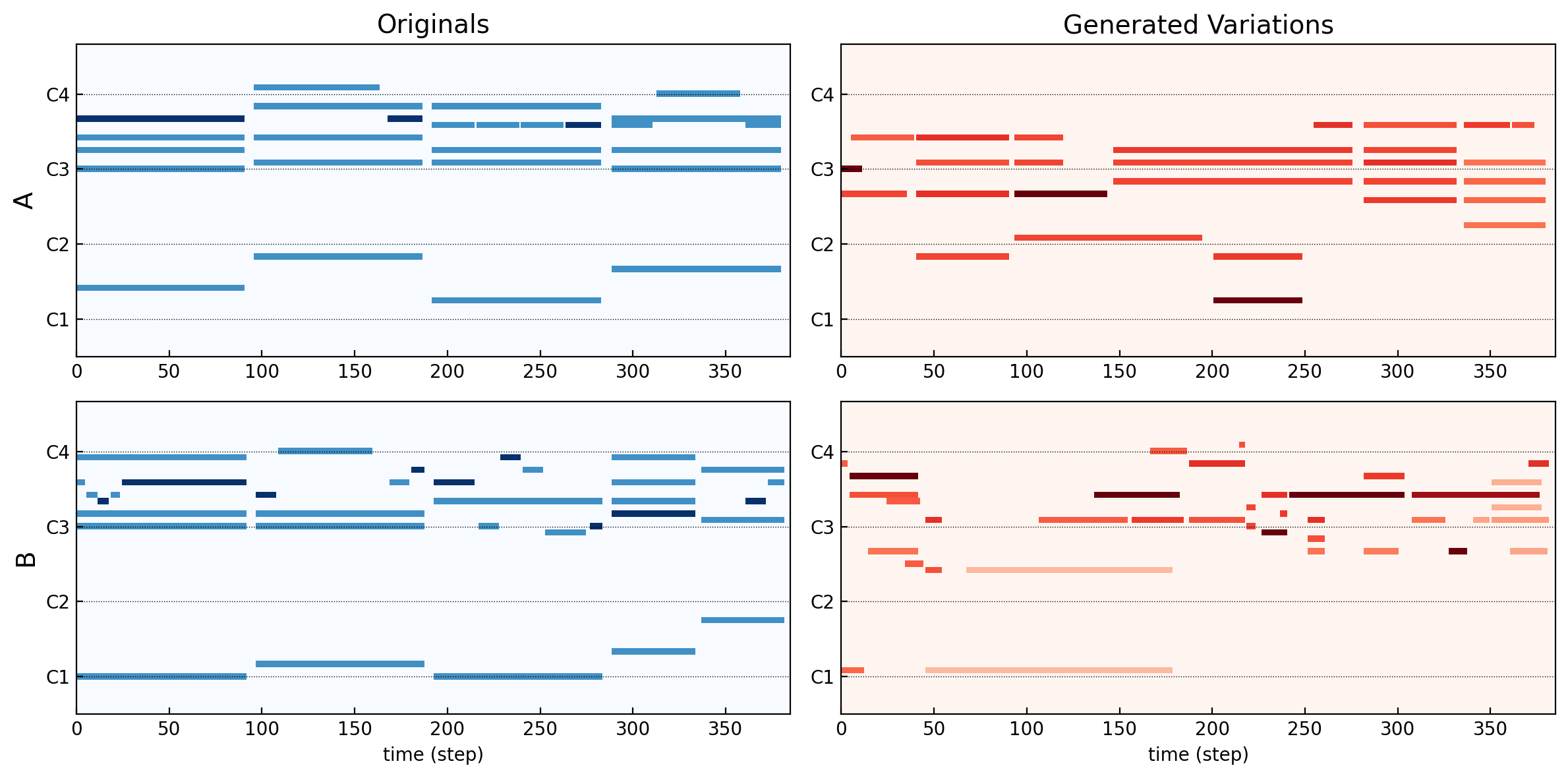}
\caption{Piano-rolls of two \textit{Original} (left in Blue) and the corresponding generated \textit{Variation} (right in Red) sections. The \textit{Original} A is from the song ``All the Things You Are'', and the \textit{Original} B is from the song ``Alfie''.}
\label{fig:pianoroll}
\end{figure}

\subsection{Results}
We present piano-rolls of two \textit{Original} segments, referred to as \textbf{A} and \textbf{B}, and the corresponding generated \textit{Variation} segments\footnote{Listening samples of the generations can be found at\ \url{https://drive.google.com/drive/folders/13SmiT2AevqP3ma3xWy4LanQwcjyRlLG1?usp=sharing}} with \textit{Original} segments used as primers to the model in Figure ~\ref{fig:pianoroll}. 
We use the same pitch-related features calculated for the dataset in Table ~\ref{tab:results} to compare the \textit{Original} segments and the corresponding generations.
According to these results, we observe that the generated \textit{Variation} segments are more complex and diverse in terms of the music features presented in Table ~\ref{tab:compare_generations}, as well as the articulation and dynamics. By listening to the generations, we find that the model's ability to accurately preserve the melody and chord patterns of the \textit{Original} segment in the generated output can be improved.  
\begin{table*}[!hbt]
\caption{Comparison of musical features for the \textit{Original} and the generated \textit{Variation} segments.} \label{tab:compare_generations}
\centering
\begin{tabular}{l@{\hskip 0.2in}cc@{\hskip 0.2in}cc}
\toprule
\textbf{Feature} & \multicolumn{2}{l}{\textbf{Original}} & \multicolumn{2}{l}{\textbf{Generated Variation}}\\
&\textbf{A}&\textbf{B}&\textbf{A}&\textbf{B}\\
\hline
Pitch Class Entropy & 2.73 & 2.75 & 2.71 & 2.86 \\
Pitch Range & 28.00 & 36.00 & 34.00 & 36.00 \\
Polyphony & 3.98 & 2.74 & 4.88 & 4.68 \\
Number of Pitches & 12.00 & 17.00 & 13.00 & 14.00 \\
Scale Consistency & 1.00 & 0.90 & 1.00& 0.98 \\
\hline
\end{tabular}
\vspace*{-3mm}
\end{table*}

\section{Conclusion} \label{future}
We present the JAZZVAR dataset a collection of 502 MIDI pairs of \textit{Variation} and \textit{Original} segments. 
We evaluated the dataset with regard to several musical features and compared the melodic and harmonic features of \textit{Variations} for different pianists performing the same \textit{Original} jazz standard. Our results indicate the diversity and complexity of \textit{Variation} in the dataset, which is one important component for successfully training a specialised generative music model. We introduced the Music Overpainting task, and trained a Music Transformer using the JAZZVAR dataset to generate \textit{Variation} segments with the \textit{Original} segments as primers.

Having a collection of \textit{Variations} performed by different pianists on the same jazz standard allows us to apply the dataset to explore tasks such as performer identification and expressive performance analysis. We aim to expand the JAZZVAR dataset in the future, using our collection of AMT MIDI data of jazz performances and corresponding jazz standards. This could either be achieved through the manual matching method as shown in Section \ref{PMP}, or through an automatic method, which would allow for a greater number of \textit{Original} and \textit{Variation} pairs to be produced. We believe that the deep generative models for the Music Overpainting task will greatly benefit from the increment of dataset size.

\printbibliography
\end{document}